\documentclass[a4paper]{raa_twocolumn}
\usepackage{natbib}
\usepackage{amssymb, amsmath, graphicx, grffile, times}
\bibpunct{(}{)}{;}{a}{}{,}
\usepackage[pagebackref=true]{hyperref}
\hypersetup{colorlinks=true, linkcolor=cyan, anchorcolor=cyan, citecolor=cyan, filecolor=cyan, urlcolor=cyan}
\footnotesize
\newdimen\digitwidth
\setbox0=\hbox{\rm0}
\digitwidth=\wd0
\catcode`!=\active
\def!{\kern\digitwidth}
\normalsize

\usepackage{threeparttable, longtable, threeparttablex, adjustbox, booktabs, url, pdflscape}

\begin{document}

\title{FAST observations of an extremely active episode of FRB 20201124A: \\
II. Energy Distribution}
\volnopage{ {\bf 20xx} Vol.\ {\bf xx} No. {\bf x}, XXX}
\setcounter{page}{1}
\author{
    Yong-Kun Zhang \inst{1,2,3},
    Pei Wang \inst{1},
    Yi Feng \inst{3},
    Bing Zhang \inst{4,5*},
    Di Li \inst{1,2,3*},
    Chao-Wei Tsai \inst{1},
    Chen-Hui Niu \inst{1},
    Rui Luo \inst{6},
    Ju-Mei Yao \inst{7},
    Wei-Wei Zhu \inst{1},
    J. L. Han \inst{1,2},
    Ke-Jia Lee \inst{8},
    De-Jiang Zhou \inst{1,2},
    Jia-Rui Niu \inst{1,2},
    Jin-Chen Jiang \inst{8},
    Wei-Yang Wang \inst{8,9},
    Chun-Feng Zhang \inst{8},
    Heng Xu \inst{8},
    Bo-Jun Wang \inst{8},
    Jiang-Wei Xu \inst{8}
}
\institute{
    National Astronomical Observatories, Chinese Academy of Sciences, Beijing 100101, China; \\
  \and
    University of Chinese Academy of Sciences, Beijing 100049, China; {\it dili@nao.cas.cn} \\
  \and
    Research Center for Intelligent Computing Platforms, Zhejiang Laboratory, Hangzhou 311100, China; \\
  \and
    Department of Physics and Astronomy, University of Nevada, Las Vegas, NV 89154, USA; {\it bing.zhang@unlv.edu} \\
  \and
    Nevada Center for Astrophysics, University of Nevada, Las Vegas, NV 89154, USA; \\
  \and
    CSIRO Space and Astronomy, PO Box 76, Epping, NSW 1710, Australia; \\
  \and
    Xinjiang Astronomical Observatory, Chinese Academy of Sciences, Urumqi, Xinjiang 830011, China; \\
  \and
    Kavli Institute for Astronomy and Astrophysics, Peking University, Beijing 100871, China; \\
  \and
    Department of Astronomy, Peking University, Beijing 100871, China; \\
  \vs \no
  {\small Received 20xx XXX; accepted 20xx XXX}
}
\abstract{
We report the properties of more than 800 bursts detected from the repeating fast radio burst (FRB) source FRB 20201124A with the Five-hundred-meter Aperture Spherical radio Telescope (FAST) during an extremely active episode on UTC September 25-28, 2021 in a series of four papers. In this second paper of the series, we study the energy distribution of 881 bursts (defined as significant signals separated by dips down to the noise level) detected in the first four days of our 19-hour observational campaign spanning 17 days. The event rate initially increased exponentially but the source activity stopped within 24 hours after the 4th day. The detection of 542 bursts in one hour during the fourth day marked the highest event rate detected from one single FRB source so far. The bursts have complex structures in the time-frequency space. We find a double-peak distribution of the waiting time, which can be modeled with two log-normal functions peaking at 51.22 ms and 10.05 s, respectively. Compared with the emission from a previous active episode of the source detected with FAST, the second distribution peak time is smaller, suggesting that this peak is defined by the activity level of the source. We calculate the isotropic energy of the bursts using both a partial bandwidth and a full bandwidth and find that the energy distribution is not significantly changed. We find that an exponentially connected broken-power-law function can fit the cumulative burst energy distribution well, with the lower and higher-energy indices being $-1.22\pm0.01$ and $-4.27\pm0.23$, respectively. Assuming a radio radiative efficiency of $\eta_r = 10^{-4}$, the total isotropic energy of the bursts released during the four days when the source was active is already $3.9\times10^{46}$ erg, exceeding $\sim 23\%$ of the available magnetar dipolar magnetic energy. This challenges the magnetar models invoking an inefficient radio emission (e.g. synchrotron maser models).
}
\authorrunning{Y. K. Zhang et al.}
\titlerunning{FRB~20201124A Energy Distribution}
\maketitle

\section{Introduction}
\label{sect:intro}

Fast Radio Bursts (FRBs) are short-duration bright transients in radio band. Since the first discovery \citep{2007Sci...318..777L}, more than 600 FRB sources have been detected \citep{2016PASA...33...45P, 2021ApJS..257...59C}. While most of the FRBs were detected once, more than 20 sources have been observed to emit repeated bursts \footnote{https://www.wis-tns.org/}. Whether FRBs can be divided into repeaters and non-repeaters is still an open question. The engine to power FRB sources and the physical mechanism to generate coherent radio emission are subject to intense study \citep{zhang20}.

Emitted energy is a fundamental property of FRBs. For the FRB population, the luminosity function is usually assumed to be a power law function with an exponential high-energy cutoff (also called Schechter function \citealt{1976ApJ...203..297S}). In the literature, the power law index $\alpha$ of the Schechter luminosity function ($dN/dL$) has been constrained by different groups using different FRB samples \citep{2016MNRAS.461L.122L,2019MNRAS.483L..93L,2018MNRAS.481.2320L, 2018MNRAS.475.1427B, 2018MNRAS.474.1900M, 2019MNRAS.489.4001G, 2020MNRAS.494..665L, 2021MNRAS.501..157Z, 2022MNRAS.511.1961H}, with $\alpha$ constrained to the range from -1.5 to -2.2. \cite{lu20} showed that an index $\sim -1.8$ can account for FRBs down to the low-luminosity Galactic FRB 200418. The observed FRB population is a convolution of the luminosity function and redshift distribution. For sources in Euclidean geometry with constant space density, the observed fluence distribution should follow a power-law distribution with an index $-1.5$ \citep{2016ApJ...830...75V}. As cosmological sources, FRBs spatial distribution deviates from this simple law and the unknown redshift distribution may play a role in shaping the observed population. If the progenitors of FRBs are young, the number density of FRBs should track the cosmic star formation history \citep{2014ARA&A..52..415M, 2017ApJ...843...84N} towards redshift about $z\sim2$. Some studies have shown that the FRB samples show the trend of deviating from the star formation history \citep{2020MNRAS.498.3927H, 2021MNRAS.501..157Z, 2021MNRAS.501.5319A, 2022MNRAS.510L..18J}. In particular, the CHIME first catalog sample seems to suggest a redshift distribution that is delayed from the star formation history \citep{2022ApJ...924L..14Z,2022MNRAS.511.1961H}, suggesting that a good fraction of FRB progenitors are likely old, with the number density decrease towards high redshift. In any case, various studies showed that the derived luminosity/energy functions are consistent with each other regardless of the assumption of redshift distribution \citep{2016MNRAS.461L.122L,2019MNRAS.483L..93L,2018MNRAS.481.2320L, 2018MNRAS.475.1427B, 2018MNRAS.474.1900M, 2019MNRAS.489.4001G, 2020MNRAS.494..665L, 2021MNRAS.501..157Z, 2022ApJ...924L..14Z, 2022MNRAS.511.1961H}.

For repeating FRBs, the energy/luminosity functions of individual sources can been studied in great detail. Detailed studies on the two well-known sources FRB~20121102A and FRB~20180916B suggested that they can be modeled by a power law function with the power-law indices ranging from $-2.8$ to $-1.7$ for bright bursts \citep{2017ApJ...850...76L, 2019ApJ...882..108W, 2019ApJ...877L..19G, 2021MNRAS.500..448C, 2021ApJ...922..115A, 2020Natur.582..351C}. These results can imply some differences for the two kinds of FRBs in respect to luminosity/energy functions \citep{2020MNRAS.494.2886H}. Owing to the high sensitivity of the Five-hundred-meter Aperture Spherical radio Telescope (FAST, \citealt{nan11}), \cite{2021Natur.598..267L} revealed a detailed energy function using more than 1600 bursts for the repeater FRB~20121102A. They found that the energy function cannot be fitted with one single power law, but requires a two-component model to fit the data. Besides the rough power law component in the high-energy end, there exists a log-normal low-energy component with characteristic values at $\sim 4.8 \times 10^{37}$ erg. It is unclear whether such a feature is universal among repeaters.

FRB~20201124A was first discovered by the Canadian Hydrogen Intensity Mapping Experiment Fast Radio Burst (CHIME/FRB) on 2020 November 24 and entered a high-active period in March 2021 \citep{2021ATel14497....1C}. Hundreds of bursts were reported during this period with a variety of radio telescopes \citep{2021MNRAS.508.5354H, 2021MNRAS.tmp.2782M, 2022ApJ...927...59L, 2022MNRAS.512.3400K, 2021ATel14526....1L, 2021arXiv211111764X}. In particular, our previous FAST observations \citep{2021arXiv211111764X} detected 1863 bursts that revealed significant short-term evolution of the rotation measure (RM). Multiple independent follow-up observations with radio interferometers localized this source to a host galaxy, which has a spectroscopic redshift of $z=0.0979$, corresponding to a luminosity distance of $451$ Mpc with $h=0.7$, $\Omega_\Lambda=0.7$ and $\Omega_m=0.3$ \citep{2021ApJ...919L..23F, 2022MNRAS.513..982R, 2021A&A...656L..15P, 2022ApJ...927L...3N}. Follow-up observations using the Keck optical telescope revealed that the host galaxy is a Milky-Way-like, metal rich, barred spiral galaxy with the FRB source residing in a low stellar density, interarm region at an intermediate galactocentric distance \citep{2021arXiv211111764X}. This source also shows its diversity in the burst properties, such as extremely high brightness and complex structure \citep{2022MNRAS.512.3400K, 2021MNRAS.tmp.2782M}, high degree of circular polarization and variation in the polarization position angle \citep{2021MNRAS.508.5354H, 2022MNRAS.512.3400K,2021arXiv211111764X}, and obvious scintillation \citep{2021MNRAS.tmp.2928M}. This previous active phase lasted until the end of May, 2021.

CHIME/FRB reported a new burst from FRB20201124A on September 21, 2021\footnote{https://www.chime-frb.ca/repeaters}. Triggered by this event, we started to use FAST to monitor this source since September 25, 2021. The observations lasted until October 17 for a total of 17 days, leading to the detection of more than 700 bursts during the first four days. In this paper, we mainly focus on the energy distribution of FRB~20201124A. The morphology, polarization and periodicity search using the same set of data are discussed in paper I (Zhou et al. 2022), III (Jiang et al. 2022) and IV (Niu et al. 2022), respectively. Section \ref{sect:data} describes our observations, burst detection and data reduction procedures. Section \ref{sect:plot} describes the results of our analysis on the burst energies and the energetics of this source. Section \ref{sect:discussion} discusses the comparison of the luminosity/energy function between this source and other sources. In Section \ref{sect:summary}, we present our conclusions.

\section{Observations and Data Processing}
\label{sect:data}

The observations of FRB~20201124A were carried out with the FAST's L-band receiver Array of 19-beams (FLAN, \citealt{FLAN, 2020RAA....20...64J}, covering a frequency bandwidth of $1-1.5$ GHz, from 2021 September 25 to October 17 with a total of 19 hours of the observation time. 

\subsection{Burst Detection}
The data recorded by FAST are in \textsc{fits} format with 4096 frequency channels and a time resolution of 49.152 $\mu s$. We first perform de-dispersion processing on the time-frequency data according to different DM values to construct new time-DM data. We then use the trained object detection model to identify the arrival time and DM of the bursts in the data. According to the DM value and burst arrival time, the data of pulse candidates are cut from the original data, and a binary classification neural network is used to determine whether they are true bursts or false signals. We also searched for bursts in dedispersed time series using a matched filtering algorithm as implemented by the \textsc{presto} program \citep{2001PhDT.......123R}. We combined the bursts detected by the two search methods as the final sample.

\subsection{DM Optimization}
Since DM is the integral of electron density along the path, we assume that the DM does not change significantly within an hour. We first use a uniform DM value of $413\ \rm pc/cm^3$ to select the pulses detected in one day from the data. Under 8 times down-sampling of time, bursts with a signal-to-noise ratio greater than 10 are selected. We then change the DM trials ranging from $-5\ \rm pc/cm^3$ to $5\ \rm pc/cm^3$ from the central value for de-dispersion to maximize the structure of all the burst profiles, and reapply this DM value to all bursts on this day.

\subsection{Flux Calibration}\label{sec:flux}
For each burst, we estimate the peak flux density with the radiometer equation
\begin{equation}\label{eq:s}
    S_\nu = \frac{{\rm SNR}\times T_{sys}}{G\sqrt{n_p\times BW \times t_{samp}}},
\end{equation}
where $T_{sys}$ is the system temperature and $G$ is the gain of the telescope, which is modeled as a function of the zenith angle and observation frequency in \cite{2020RAA....20...64J}, $n_p=2$ is the number of polarization summed, $t_{samp}$ is the sampling time, $BW$ is the used bandwidth for calibration. 

Repeating FRBs tend to have a narrow emission bandwidth compared with apparent non-repeating events \citep{2021ApJ...923....1P}. The bandwidth may be related to the telescope sensitivity. For these narrow-band bursts, usually the energy is calculated using the flux density defined by the full bandwidth to be multiplied by the full bandwidth. Alternatively, the energy can be calculated using the flux defined with the narrow band in which the burst is detected and multiplied by this narrow band. We roughly estimate the emission bandwidth of each burst by dividing the whole bandpass into multiple 50-MHz sub-bands and identifying the sub-bands containing burst emission. If the calculations are performed correctly, the two methods should result in the same results. In order to check this, we calculate $S_\nu$ using two ways. The first is to assume that the burst energy is spread in the full bandwidth of the telescope $BW=500\ {\rm MHz}$ for all the bursts. The second is to use the true $BW$ for each individual burst. When different bandwidths are adopted, the noise level and the signal-to-noise ratio of a pulse change. Therefore, when using the observed pulse bandwidth for flux calibration, we recalculated the signal-to-noise ratio and the $T_{sys}$ and $G$ of the corresponding frequency bandwidth.

\section{Results}
\label{sect:plot}

\subsection{Rate and Waiting Time}

\begin{figure}[!htp]
    \centering
    \includegraphics[width=0.45\textwidth]{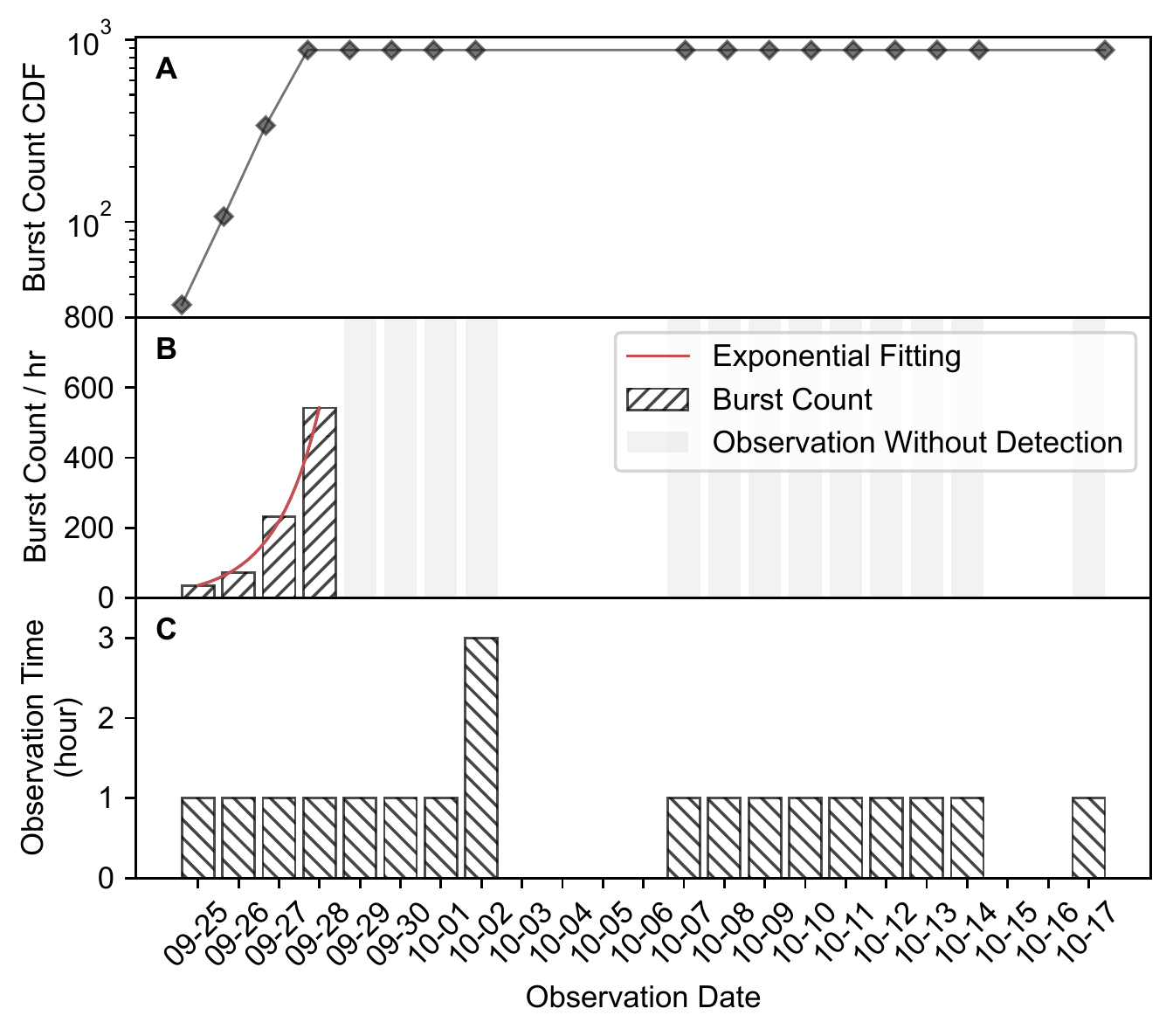}
    \caption{Observation and Detection of FRB~20201124A. A: the cumulative distribution of burst detection. B: the burst count in each observation day. Gray bars are the observation sessions without bursts detection, the red line shows the exponential fitting of the burst detection. C: the length of each observation session.}
    \label{fig:burstcount}
\end{figure}

During the 17-day observation campaign, we detected 35, 72, 232, 542 bursts from FRB~20201124A in the first four days (i.e. from September 25 to September 28), for a total of number of 881 bursts. The bursts are defined as significant signals separated by dips down to the noise level in this paper\footnote{We note that the definitions of a burst are different among the four papers in this series, because different groups adopted different criteria for the different scientific purposes in the respective papers.}. The observation and detection are shown in Figure~\ref{fig:burstcount}. The cumulative distribution is plotted in the linear-logarithmic scale. The linear line in the figure suggests that the accumulated number increases exponentially with time. Interestingly, the bursts suddenly disappeared in the next day, which is unexpected. The peak burst rate on the third and fourth day reached $232/ \rm hr$ and $542/ \rm hr$, respectively, which exceeded all previously reported burst rates from this and other repeating FRBs, including $122/ \rm hr$ for FRB~20121102A \citep{2021Natur.598..267L}, $4.5/ \rm hr$ for FRB~20190520B \citep{2022Natur.606..873N}, and $45.8/ \rm hr$ for FRB~20201124A itself during the previous active session \citep{2021arXiv211111764X}.

\begin{figure}[!htp]
    \centering
    \includegraphics[width=0.45\textwidth]{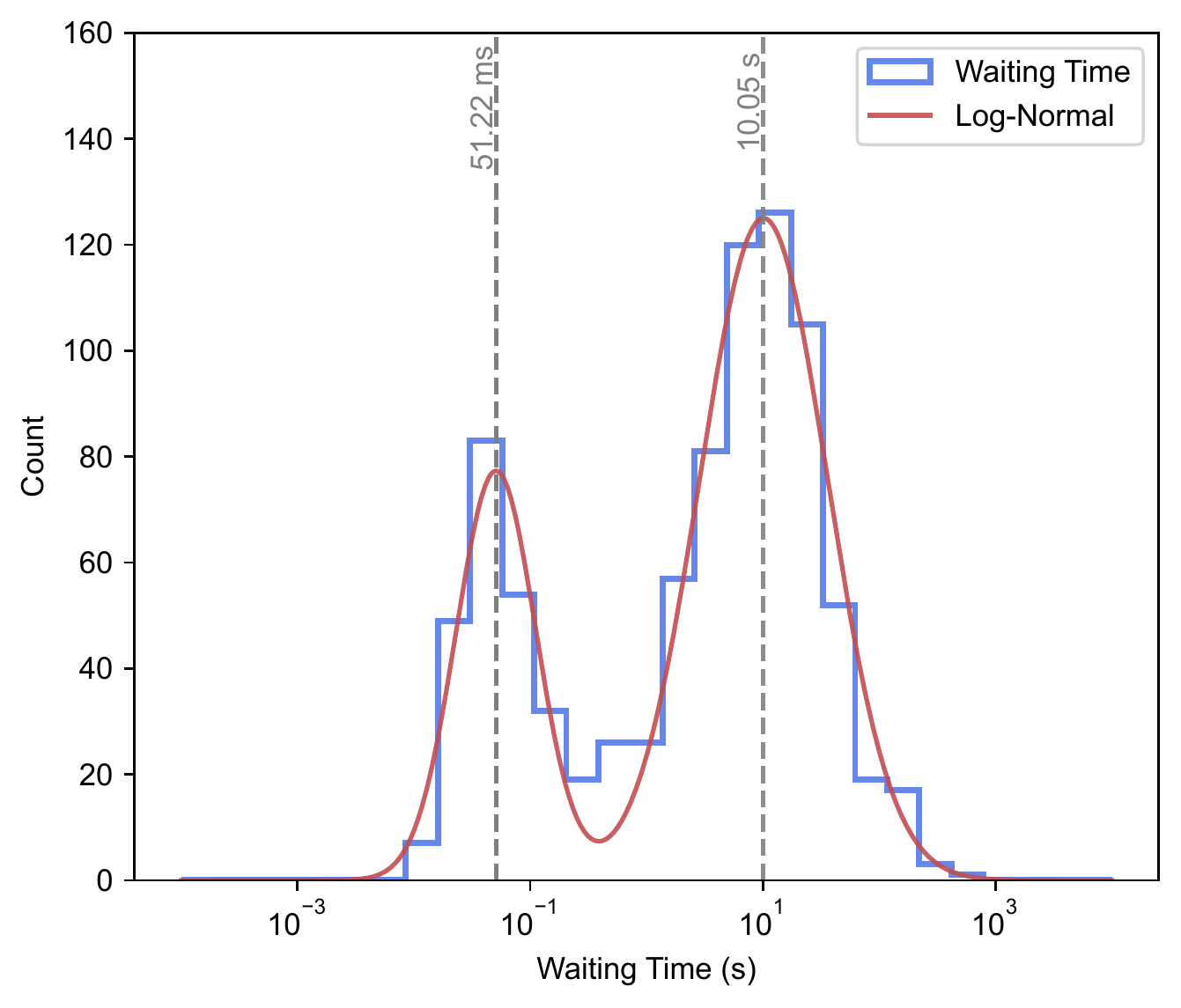}
    \caption{Waiting time distribution of FRB~20201124A. The blue step and red line show the distribution of waiting time and two log-normal fitting.}
    \label{fig:watime}
\end{figure}

Waiting time is defined as the difference between the arrival time of two adjacent bursts. All the waiting times were calculated in the same observation session to avoid the observation gap of about 23 hr. The waiting time distribution of FRB~20201124A is shown in Figure~\ref{fig:watime}, which can be well-fitted by two log-normal functions peaking at $10.05$ s and $51.22$ ms, respectively. Note that the waiting time distribution of FRB~20201124A in the previous active period was also a bimodal distribution, but the characteristic timescale of the second (long-duration) peak was $135$ s \citep{2021arXiv211111764X}. Our observation suggests that this characteristic waiting time is not universal within one source, but rather depends on the activity level of the source. 
The bimodal form of waiting time has also been found in FRB~20121102A \citep{2021Natur.598..267L,2022MNRAS.515.3577H, 2021ApJ...922..115A}. The timescale of the second peak varies also significantly for telescopes with different sensitivities, with shorter waiting times for more sensitive telescopes (which detect more bursts). 

\subsection{Energy calculation}
We calculate the isotropic equivalent burst energy following the Equation:
\begin{equation}\label{eq:g}
    E = 10^{39} {\rm erg}\frac{4\pi}{1+z} \left(\frac{D_L}{10^{28}{\rm cm}}\right)^{2} \left(\frac{F_{\nu}}{\rm Jy\cdot ms}\right)\left(\frac{\Delta\nu}{\rm GHz}\right),
\end{equation}
where $D_L=453.3 {\rm Mpc}$ is the luminosity distance of FRB~20201124A corresponding to the redshift $z=0.09795$ acquired in \cite{2021arXiv211111764X}, $F_\nu=S_\nu\times W_{\rm eq}$ is the specific fluence, $S_\nu$ is the peak specific flux, $\Delta \nu$ is the bandwidth of each pulse. The adoption of $\Delta \nu$ is more relevant for narrow-band FRBs, which is typically the case for the bursts in repeaters. As discussed in \S\,\ref{sec:flux}, we adopt two choices of $BW$ to calculate $S_\nu$ and compare the derived $E$ values.

\begin{figure}[!htp]
    \centering
    \includegraphics[width=0.45\textwidth]{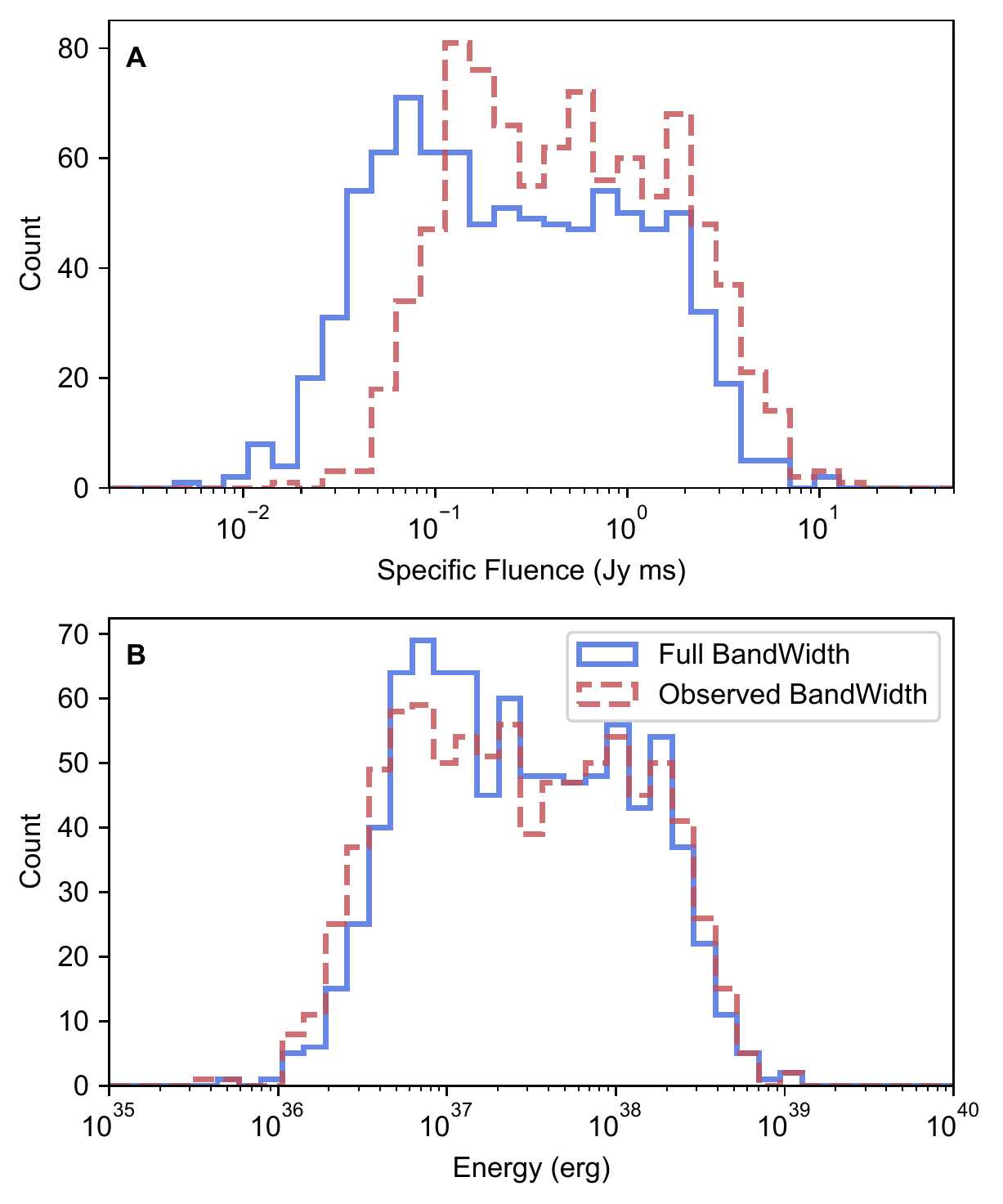}
    \caption{Specific fluence and energy distribution. Energy is calculated by Eq.~\ref{eq:g}. Solid blue lines are distribution for full bandwidth, i.e. $\Delta\nu=500 {\rm MHz}$, red dashed lines are for observed burst bandwidth.}
    \label{fig:fe}
\end{figure}

Figure~\ref{fig:fe}A shows the specific fluence ($F_\nu$) and the energy distribution of bursts calculated with $\Delta\nu=500 {\rm MHz}$ and the observed bandwidth for individual bursts, respectively. The specific fluence derived using the observed narrow bandwidth is greater than that derived assuming the full bandwidth. This is natural because the definition of $F_\nu$ is the average over frequency channels. Since the energy contained in the observed bandwidth is similar to that in the full bandwidth but the observed bandwidth is narrower than the full bandwidth, the average value for the full bandwidth is smaller. Figure~\ref{fig:fe}B presents the energy distribution of the two methods after multiplying by the respective bandwidth. It can be seen that the derived energy distributions between the two methods appear to be similar to each other, as is expected. The p-value from the K-S test of the energy distribution calculated with two types of bandwidth is 0.35, which cannot reject the hypothesis that the two energy distributions are consistent with each other. Therefore, we conclude that the energy calculation is not significantly affected by the adoption of the bandwidth, and the normal methods using the full bandwidth to calculate the energy is reliable. 

\cite{2021Natur.598..267L} introduced the detection of 1652 pulses from FRB~20121102A with FAST and proposes a bimodal energy distribution. The calculation there following Eq.~9 from  \cite{2018ApJ...867L..21Z}, which differs from Eq.~\ref{eq:de} by a constant $\nu_c/\Delta\nu$, if we calibrate the fluence both with the full bandwidth\footnote{The method of \cite{2018ApJ...867L..21Z} applies to FRBs with emission bandwidth much wider than the telescope's bandwidth, which seems to be the case for at least some non-repeating FRBs.}. 
\cite{2021ApJ...920L..18A} claimed that he recalculated the energy using the bandwidth in Table 1 of \cite{2021Natur.598..267L}, and the resulting distribution did not show a significant bimodality. As we have shown in Figure~\ref{fig:fe}A, the specific fluence varies greatly with the bandwidth selection. The calculation in \cite{2021ApJ...920L..18A} simply replaced the $\nu_c$ with the apparent observed bandwidth for individual bursts, but still used the fluence values derived from the full bandwidth to do the calculation. Therefore the calculations of \cite{2021ApJ...920L..18A} resulted in incorrect results. The bimodal energy distribution for FRB 20121102A as claimed by \cite{2021Natur.598..267L} remains intact.

\subsection{Energy Distribution}

Energy/luminosity function can provide important hints on the emission mechanism of the source. As we demonstrated in the previous subsection, the use of limited bandwidth or full bandwidth does not change the energy distribution. Therefore, we uniformly use Eq.~\ref{eq:g} and the full bandwidth fluence for energy calculation in the following analysis.

\begin{figure}[!htp]
    \centering
    \includegraphics[width=0.45\textwidth]{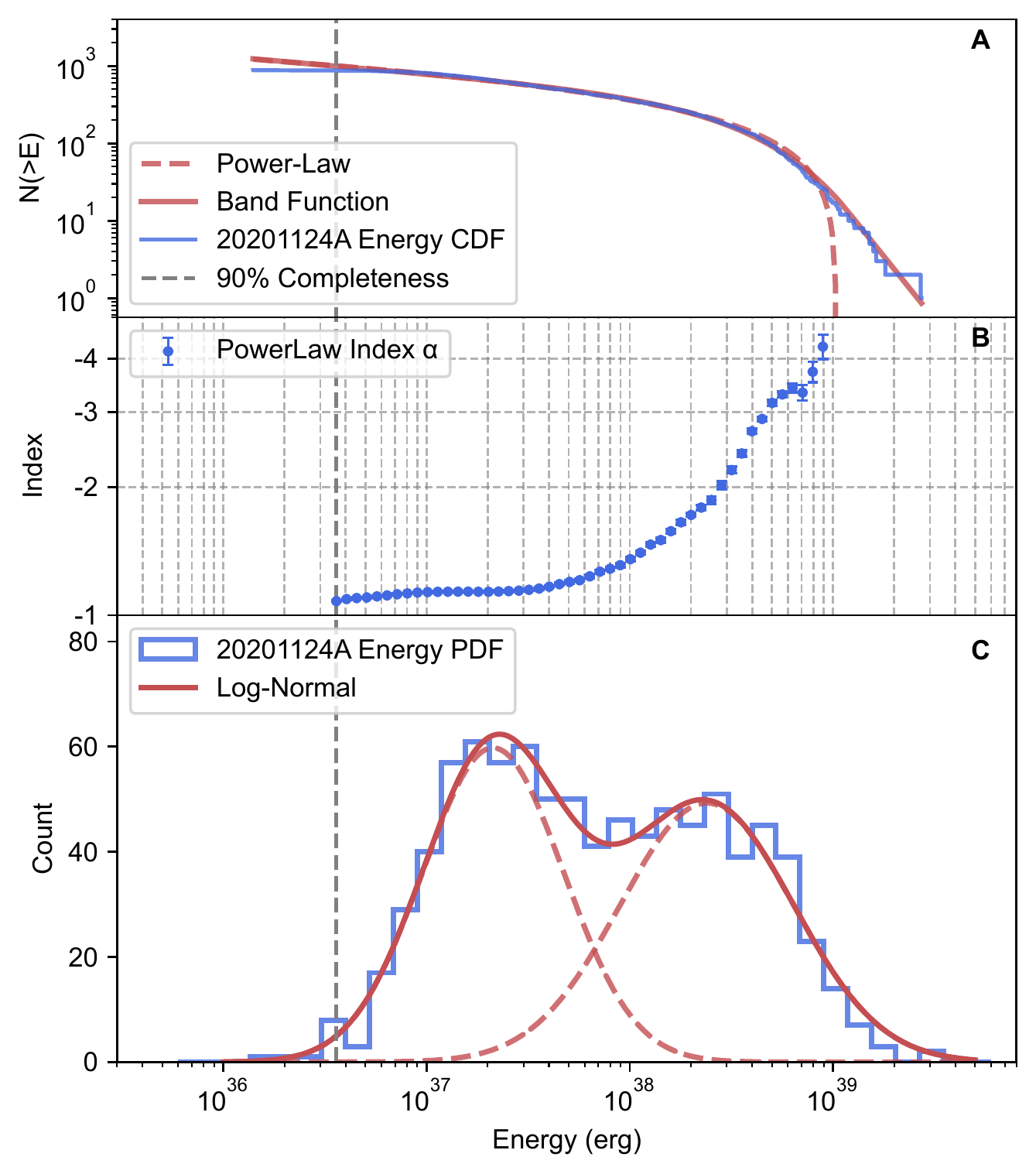}
    \caption{The energy distribution of FRB~20201124A. \textbf{A}: the cumulative energy distribution of FRB~20201124A (blue) with Band function and single-power-law function fitting. \textbf{B}: the index of single-power-law fitting as the function of energy threshold. \textbf{C}: the energy distribution of FRB~20201124A (blue) with two log-normal fitting.}
    \label{fig:ef}
\end{figure}

As shown in Figure~\ref{fig:ef}C, the energy distribution of FRB~20201124A also presents a clear bimodal distribution, which can be well-fitted by two log-normal functions peaking at $2.27\times10^{37}$ erg and $2.28\times10^{38}$ erg, respectively. This may indicate the existence of more than one emission mechanism to generate FRBs. The distribution of FRB~20121102A could be also  fitted by a similar double log-normal distribution (Extended Data Fig. 5 in \citealt{2021Natur.598..267L}), but the peak energies is not the same for the two FRB sources.

For a power law energy/luminosity distribution $dN/dE = A E^\alpha$, the cumulative energy distribution can be calculated as
\begin{equation}\label{eq:de}
    \begin{aligned}
    N(>E)&=\int_E^{E_{\rm max}} dN = \int_E^{E_{\rm max}} AE^{\alpha} dE\\
    &= \frac{A}{-\alpha-1}\left[E^{\alpha+1}-E_{\rm max}^{\alpha+1}\right]=\frac{A}{-\hat\alpha}\left[E^{\hat\alpha}-E_{\rm max}^{\hat\alpha}\right],
    \end{aligned}
\end{equation}
where $E_{\rm max}$ is the maximum energy for a certain sample, and $\hat\alpha=\alpha+1$. One can see that in general, the cumulative energy distribution is also roughly a power law with index $\hat\alpha$ is $\hat\alpha < 0$. Different types of astrophysical sources have different $\hat\alpha$ values. For example, solar flare events have $-0.9 < \hat\alpha < -0.5$ \citep{2015EP&S...67...59M} and giant pulses of the Crab pulsar have $-2.8<\hat\alpha<-1.1$  \citep{2012ApJ...760...64M, 2021FrPhy..1624503L}. Different repeating FRBs are observed to have different $\hat\alpha$ values: e.g. $-1.8 < \hat\alpha < -0.7$ for FRB~20121102A \citep{2017ApJ...850...76L, 2021MNRAS.500..448C, 2019ApJ...877L..19G}, and $\hat\alpha \sim -1.3$ for FRB~20180916B \citep{2020Natur.582..351C}. For FRB~20201124A, $\hat\alpha \sim -1.2$ was measured for the uGMRT bursts \citep{2021MNRAS.tmp.2782M} and $\hat\alpha \sim -3.6$ was measured for the CHIME bursts \citep{2022ApJ...927...59L}. A broken power-law was also used to model the cumulative energy distribution of FRB~20121102A, the index $\hat\alpha = -1.4$ and $\hat\beta = -1.8$ around the turning point $2.3\times10^{37}$ erg \citep{2022MNRAS.515.3577H, 2021ApJ...922..115A}. The bursts in our FAST sample also require a two-segment power law, but the turning point is not obvious. Therefore, we apply an exponentially connected broken-power-law function (which is called the Band function proposed by \cite{1993ApJ...413..281B} to describe the gamma-ray spectra of gamma-ray bursts), which reads
\begin{equation}\label{eq:ce}
    N(>E)=\left\{\begin{aligned}
        & AE^{\hat\alpha}\, e^{\left(-E/E_0\right)} \quad &E\le(\hat\alpha-\hat\beta)E_0\\
        & AE^{\hat\beta} \, \left[\frac{(\hat\alpha-\hat\beta)E_0}{e}\right]^{\hat\alpha-\hat\beta} \quad &E\ge(\hat\alpha-\hat\beta)E_0\\
    \end{aligned}\right.
\end{equation}
where $\hat\alpha=\alpha+1$ and $\hat\beta=\beta+1$ are the power-law indices of the lower and higher cumulative energy distributions, and $E_0$ is a characteristic energy. Note that strictly speaking the cumulative energy distribution of two log-normal functions should have a complicated shape. The simple power law function is insufficient. We therefore adopt the next simplest function, i.e. a smoothly joint two-segment broken power law function (the Band function) to fit it. As shown in Figure~\ref{fig:ef}A, this function fits the data reasonably well. In principle, the turnover in the low-energy end of the low-energy log-normal component would demand an additional break in the function to fit the cumulative distribution and indeed there is a deviation at the low-energy end between the data and fitting function. In any case, when considering possible selection effect near the sensitivity threshold (which would compensate the deficit of low-energy bursts from the data), we believe that the Band function presents a reasonable description of the cumulative distribution function.


We exclude the bursts that are below the 90\% detection threshold ($E_{\rm th}=3.6\times10^{36}$ erg), which makes up 1\% of the total sample. We then randomly select 90\% of the bursts in the sample to generate a cumulative energy distribution. We repeat this process 1000 times and fit these 1000 cumulative energy distributions with a power-law function and the Band function. The central limit theorem makes the fitting parameters converge to a Gaussian distribution. We use the mean and standard deviation of the Gaussian distribution as the standard value and error of the fitting parameters. Figure~\ref{fig:ef}A shows the fitting results of the cumulative energy distribution. The power-law index is $\hat\alpha=-0.08\pm0.01$. The index values in the Band function are $\hat\alpha=-0.22\pm0.01$ and $\hat\beta=-3.27\pm0.34$, with the turning point at $(1.1\pm0.2)\times10^{39}$ erg. Note that the power-law indices and the break energy are all different from those derived from the CHIME bursts. Considering the uncertainties, the index of higher energy end is consistent with the CHIME result.

We repeat the fitting process with a varying energy threshold. The index $\hat\alpha$ of a single-power-law function does not converge to a single value in panel B of Figure~\ref{fig:ef}. This is another indication that a single power-law distribution is inadequate to describe the cumulative energy distribution of this FRB source.

\section{Discussion}
\label{sect:discussion}

\subsection{Comparison with other repeating FRBs}

\begin{figure}[!htp]
    \centering
    \includegraphics[width=0.45\textwidth]{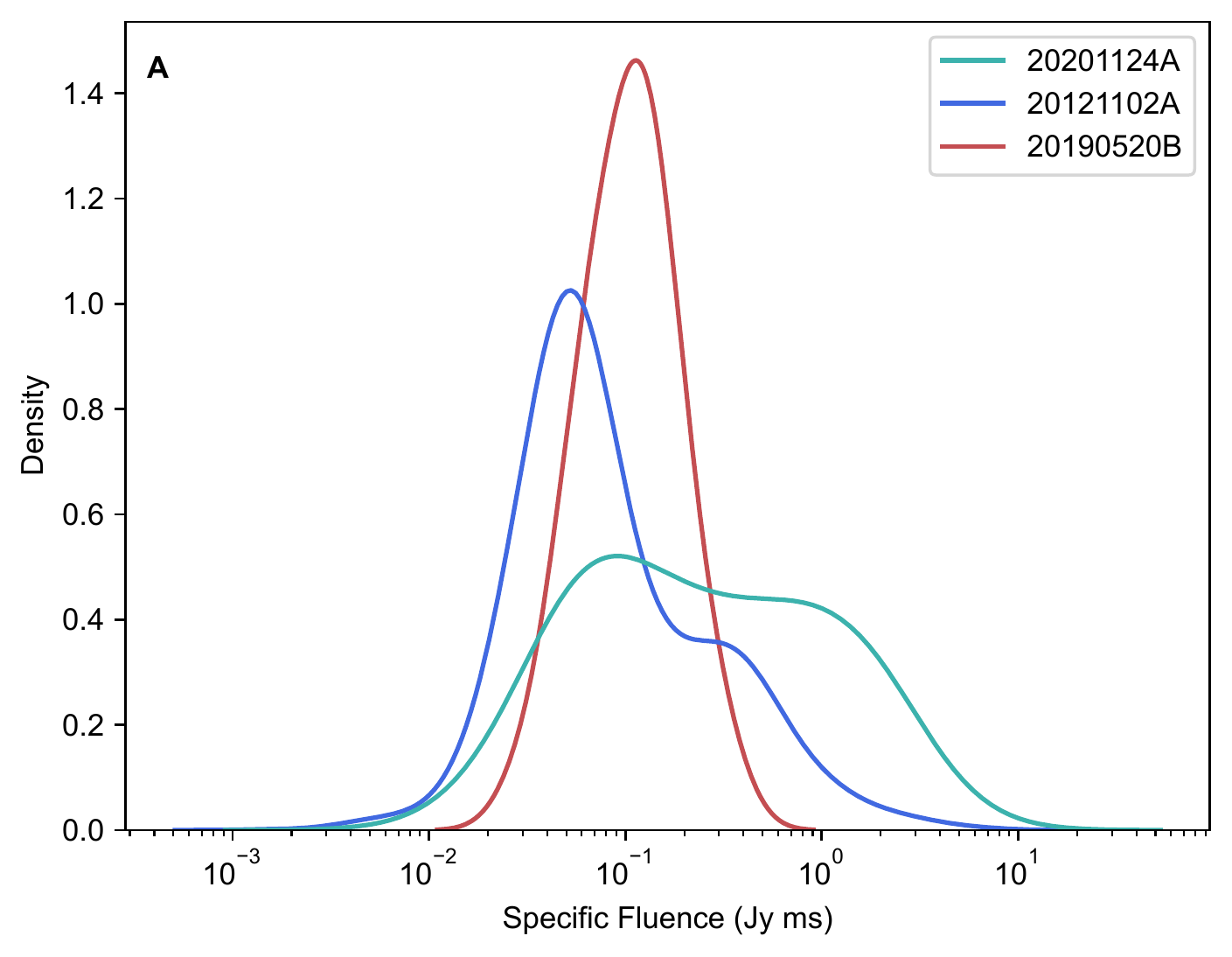}\\
    \includegraphics[width=0.45\textwidth]{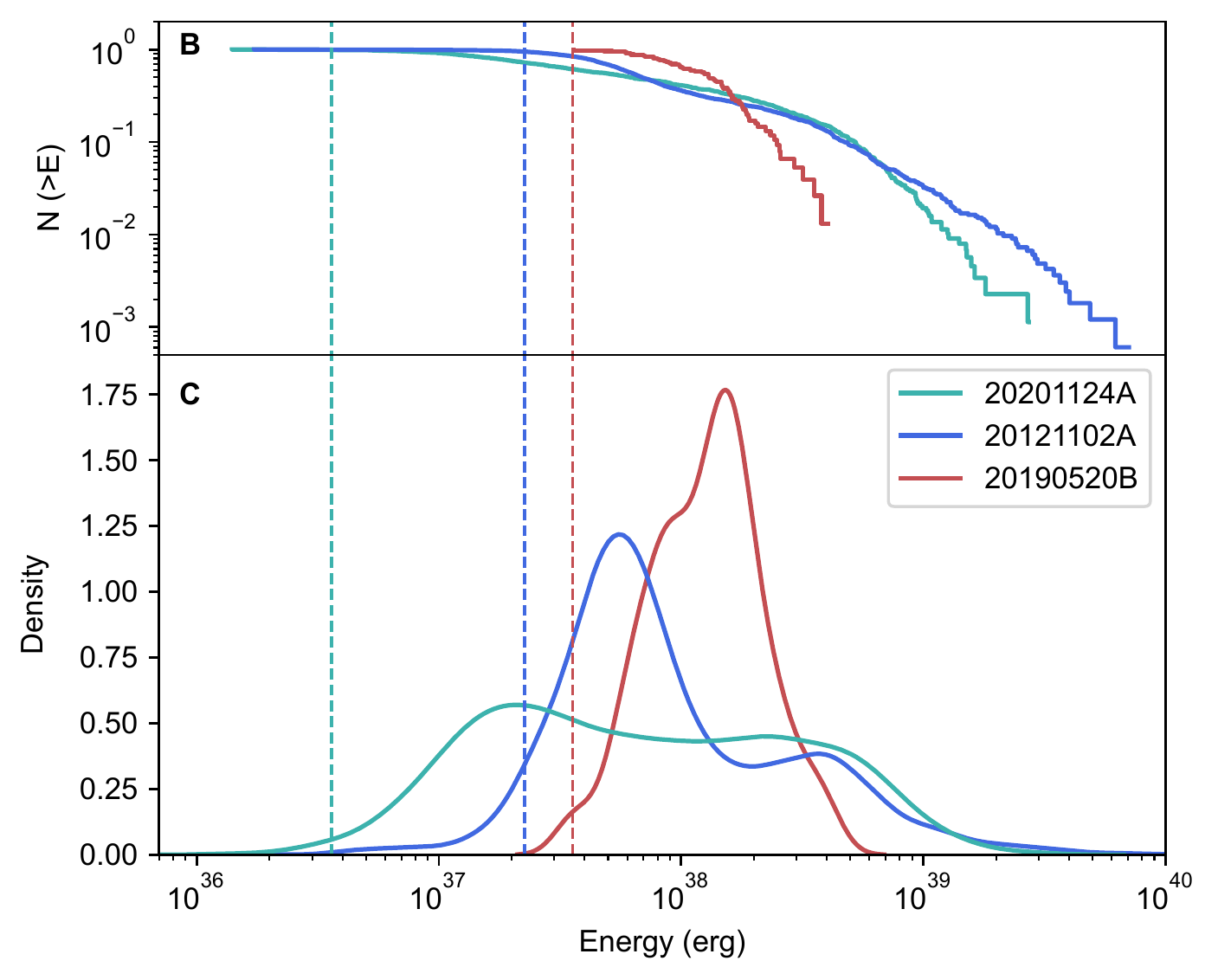}
    \caption{Fluence and energy distribution of FRB~20201124A (green), FRB~20190520B (red) and FRB~20121102A (blue). \textbf{A} is the kernel density estimation (KDE) of the three FRBs' fluence. \textbf{B} and \textbf{C} are cumulative distributions and KDEs of the three FRBs' energy. The dashed lines are the detection thresholds.}
    \label{fig:fc}
\end{figure}

FRB~20121102A, FRB~20190520B and FRB~20201124A are three repeating FRBs which have been extensively-studied by FAST. FRB~20121102A is the first known repeating FRB \citep{2014ApJ...790..101S, 2016Natur.531..202S}. FRB~20190520B is a repeating FRB detected by Commensal Radio Astronomy FAST Survey (CRAFTS) \citep{Li_2019, 2022Natur.606..873N}. Both FRB~20121102A and FRB~20190520B are found to be associated with a compact persistent radio source (PRS) \citep{2017Natur.541...58C, 2017ApJ...834L...8M, 2022Natur.606..873N}. FRB~20201124A is also associated with persistent radio emission, but it is not a point source and is rather extended in space \citep{2022MNRAS.513..982R, piro21}. The FAST-measured energy distributions of these three FRB sources can be directly compared against each other to reduce the possible selection biases caused by using different telescope systems. Figure~\ref{fig:fc} shows the energy distributions of the three FRBs in the form of both power density distribution (PDF) and cumulative density distribution (CDF). Both the PDF and CDF show that the energy distributions of the three FRB sources are not consistent.

Table~\ref{tab:energycal} shows the observed isotropic radio energy and the inferred burst energy of the three FRBs. The total observed energy $E_{\rm radio}$ is the sum of all the observed burst energies in the radio band, which is corrected to the total energy released during the bursts via Eq.\ref{eq:re}.
\begin{equation}\label{eq:re}
    E_{\rm bursts} = E_{\rm radio}\times F_b\times \eta^{-1} \times \zeta^{-1},
\end{equation}
where $E_{\rm bursts}$ is the total energy emitted by FRB source during bursts, $F_b$ is the beaming factor, $\eta$ is the radio efficiency which is normalized to $\sim 10^{-5}$ (similar order to FRB 20200428), and $\zeta$ is the observation duty cycle. For example, the observations of FRB~20201124A this time were carried out in 4 sessions in 4 days with 1 hour per session. The duty cycle $\zeta$ is therefore estimated as $(4 \times1\ {\rm hour})/(4\times24\ {\rm hours}) = 1/24$.

Compared with other repeaters, FRB~20201124A has the largest averaged energy, especially on September 28. Taking the nominal values of $\eta = 10^{-5}$ and $F_b=0.1$, the total burst energy released on September 28 has reached $(2.46 \times 10^{46} \ {\rm erg}) \eta_{-5}^{-1} F_{b.-1}$. Compared with the total dipolar magnetic energy of a magnetar $E_{\rm mag} = (1/6) B_p^2 R^3 \sim (1.7\times 10^{47} \ {\rm erg}) B_{p,15}^2 R_6^3$, the burst energy emitted on this day already exceeded $14.3\% \ \eta_{-5}^{-1} F_{b,-1}$ of the available magnetar energy. This raises an even more severe energy budget problem compared with FRB~20121102A \citep{2021Natur.598..267L}. This requires that the radio emission efficiency should be quite high for the magnetar model or that the magnetar engine does not apply. For magnetar models, since the synchrotron maser models invoking relativistic shocks are quite inefficient, the data favors magnetospheric models for FRBs, which can have a higher efficiency as observed in radio pulsars \citep{2017MNRAS.468.2726K,2018ApJ...868...31Y,lu20,2022ApJ...925...53Z}. 
The synchrotron maser model may be still relevant if the central engine is not a magnetar, but rather an accreting black hole system \citep[e.g.][]{2021ApJ...917...13S}, whose total energy budget is not limited by the total magnetic energy of a magnetar.

\begin{table*}[!htp]
\centering
\setlength{\tabcolsep}{5pt}
\caption{Energy budget of three repeating FRBs.}\label{tab:energycal}

\begin{threeparttable}
\begin{small}
\begin{tabular}{cccccccc}
 \toprule
  Name         & ObDays & ObTimes              & Total Observed Energy\tnote{a} & Averaged Energy\tnote{b} & Total Energy\tnote{c}  \\
               & (day)  & $T_{\rm obs}$ (hour) & $E_{\rm radio}$ (erg)          & $\bar L_{\rm radio}$ (erg/s) & $E_{\rm bursts}$ (erg) \\
 \midrule
  FRB~20201124A & 4      & 4                    & $1.60\times10^{41}$            & $1.11\times10^{37}$          & $3.85\times10^{46}$             \\
  FRB~20201124A-0928 & 1 & 1                    & $1.02\times10^{41}$            & $2.84\times10^{37}$          & $2.46\times10^{46}$             \\
  FRB~20121102A & 47     & 59.5                 & $3.41\times10^{41}$            & $1.59\times10^{36}$          & $6.47\times10^{46}$             \\
  FRB~20190520B & 11     & 18.5                 & $1.10\times10^{40}$            & $1.65\times10^{35}$          & $1.56\times10^{45}$             \\
 \bottomrule
\end{tabular}
\end{small}
\begin{tablenotes}
    \footnotesize
    \item [a] {Sum of the observed isotropic radio energies of all bursts.}
    \item [b] {Defined as $\rm E_{\rm radio} / T_{\rm obs}$}
    \item [c] {$E_{\rm bursts} = E_{\rm radio} \times \left[(\zeta={\rm ObTimes/ObDays})^{-1} \times (\eta=10^{-5})^{-1} \times (F_b=0.1)\right]$}
\end{tablenotes}
\end{threeparttable}
\end{table*}

\subsection{Comparison of different burst definition schemes}

To quantify the specific morphological characteristics of the bursts, we differentiate two kinds of pulse profile classification schemes. The first classification scheme is based on a judgment of pulse flux threshold, taking the pulse profile above the detection threshold as a separated burst. For conservative considerations, we take the detection threshold using a 3$\sigma$ noise level of the baseline due to the complicated effect on the actual sensitivity, e.g., the bandwidth limited structure of the bursts and radio frequency interference (RFI) events. This definition is also the pulse definition used in previous work, and is the same as the definition used for FRB~20121102A \citep{2021Natur.598..267L} and FRB~20190520B \citep{2022Natur.606..873N}. In the following text, we call this definition Combined Bursts Definition (CBD).

\begin{figure}[!htp]
    \centering
    \includegraphics[width=0.45\textwidth]{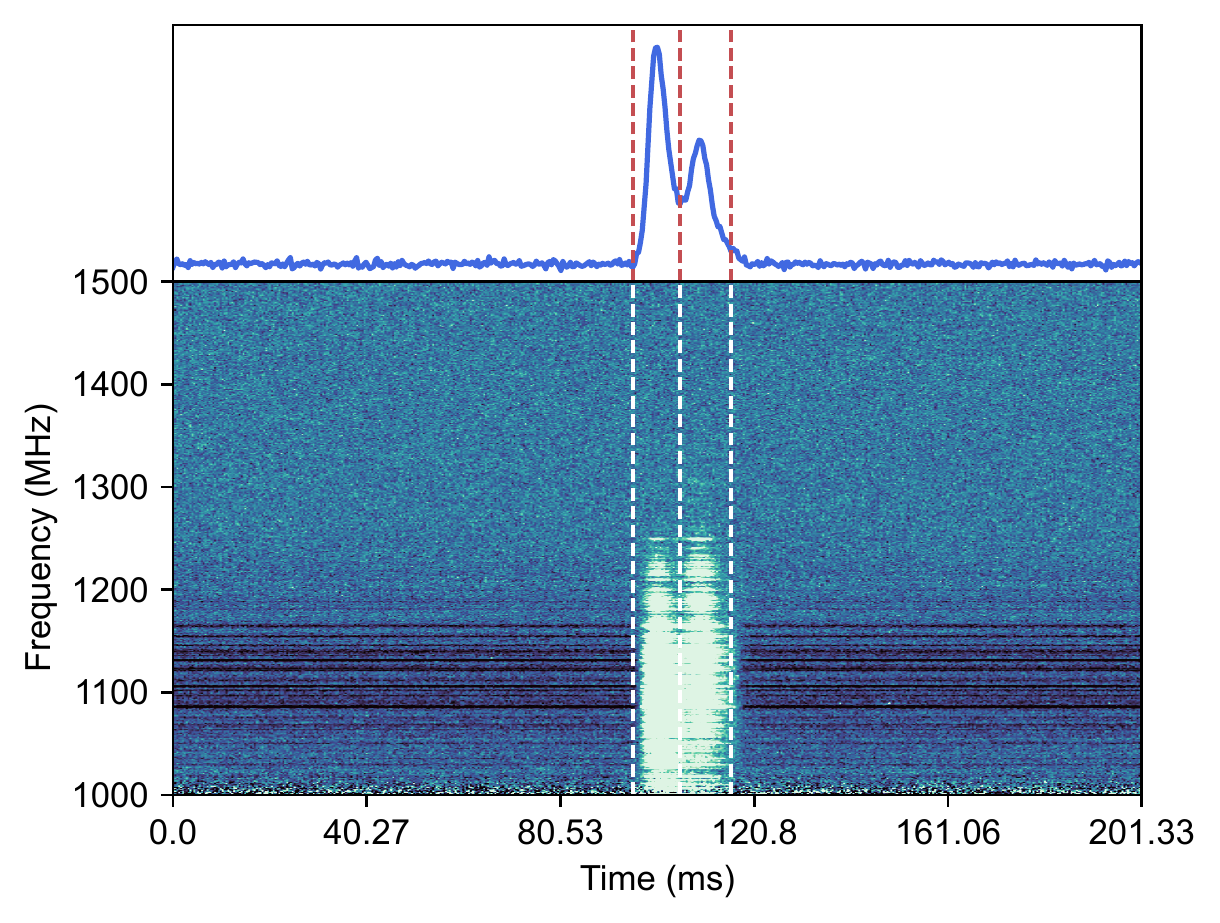}
    \caption{A burst event example detected on September 26.}
    \label{fig:burst}
\end{figure}

For another method, we try to differentiate the bursts by referring to the morphological clustering characteristics of the pulses in the dynamic spectrum, which means that pulses that can be distinguished in the time-frequency diagram are treated as different bursts, like Figure~\ref{fig:burst}. We call this definition Separated Bursts Definition (SBD). As a result, the event in Figure~\ref{fig:burst} contains 2 bursts under SBD but 1 burst under CBD. All bursts under SBD are artificially isolated and flux calibration is performed independently.

\begin{figure}[!htp]
    \centering
    \includegraphics[width=0.45\textwidth]{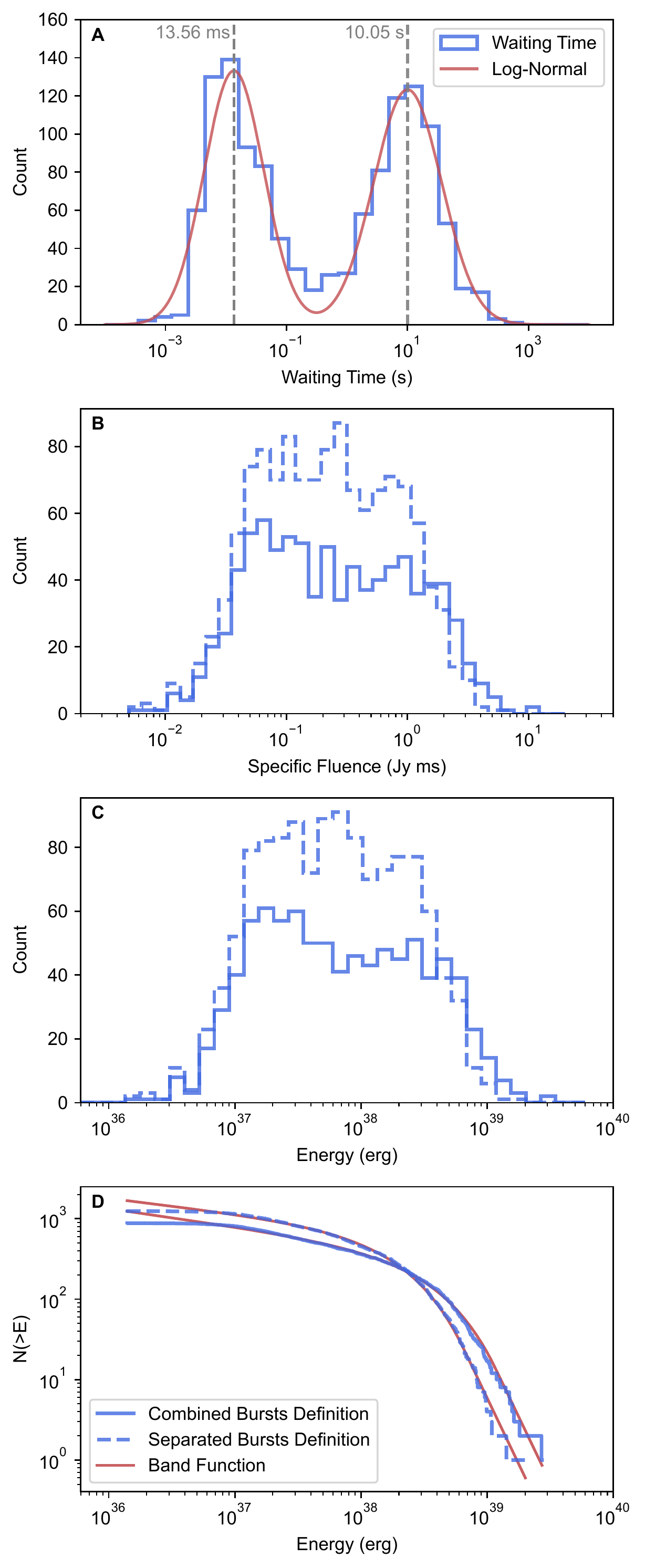}
    \caption{A: Waiting time distribution of the bursts from FRB~20201124A with SBD. The red line shows the log-normal fitting. B: Specific fluence distribution of bursts with CBD (blue solid line) and SBD (blue dashed line). C: Energy distribution of bursts with CBD and SBD. D: Cumulative energy distribution of bursts with CBD and SBD with Band function fitting (red line).}
    \label{fig:et}
\end{figure}

Figure~\ref{fig:et}A shows the waiting time distribution of bursts with SBD, which is still a bi-modal distribution and can be modeled with two log-normal functions peaking at $13.56$ ms and $10.05$ s, respectively. The second peak time around 10 s is consistent with the waiting time distribution of CBD. However, the first peak time is significantly reduced. The distributions of specific fluence and energy exhibit distinct distributions under the two burst definitions. We use the K-S test to check whether the energies calculated from the two burst definition schemes belong to the same distribution. The p-value is 0.2\%, meaning we could rule out the hypothesis that they were from the same distribution. We also use the Band function to fit the cumulative energy distribution of the bursts with SBD. The index values in the Band function are $\hat\alpha=-0.18\pm0.01$ and $\hat\beta=-3.27\pm0.16$, respectively. The index in the high energy regime is the same with CBD, while the index in the lower energy regime is slightly smaller than that of CBD. The turning point is $(6.4\pm0.1)\times10^{38}$ erg, which is smaller than that for CBD. While we still do not know the true nature of FRBs, different definitions of bursts may lead to somewhat different analysis results. We caution that when studying burst energy distributions of FRBs, one needs to specify a burst definition scheme. 

\section{Conclusions}
\label{sect:summary}
FRB~20201124A is a repeating FRB discovered by CHIME \citep{2021ATel14497....1C}. FAST has monitored source multiple times. Besides an earlier observational campaign reported by \cite{2021arXiv211111764X}, we study a very active 4-day emitting episode on September 25-28, 2021, in this series of papers. The main results of this paper are as follows.

\begin{itemize}
\setlength{\itemsep}{3pt}

    \item[1)] Using our CBD definition method, a total of 881 bursts were detected between September 25 to September 28.
    
    \item[2)] The apparent event rate increased exponentially in 4 days and then dropped to zero after Sep. 29.
    
    \item[3)] The waiting time of FRB~20201124A shows a bimodal structure, which can be well-fitted by two log-normal functions. The longer waiting times peak around 10 s and is insensitive to the burst definition criterion. It is much shorter than that of the previous epoch \citep{2021arXiv211111764X}, suggesting that it is related to the active level of the source.
    
    \item[4)] The shorter waiting times peak around either 51~ms or 14~ms, depending on the definition of burst separation (see Section 3). 
    
    \item[5)] The peak event rate of 542/hr exceeds all the detected FRBs in the past, including 122/hr of FRB~20121102A \citep{2021Natur.598..267L}, and FRB~20201124A itself, which was 45.8/hr during the last active period \citep{2021arXiv211111764X}. The total burst energy released during this epoch is a significant fraction of magnetar total energy unless the radio emission frequency is high. This challenges the synchrotron maser mechanism for FRBs and even the magnetar source model. 
   
    \item[6)] The energy distribution cannot be described by a single power-law. The cumulative distribution of energy can be well modeled by an exponentially connected broken power law known as the Band function. The slopes at low and high energy regimes are $\hat\alpha=-0.22\pm0.01$ and $\hat\beta=-3.27\pm0.23$, respectively.
    
    \item[7)] The energy distributions of three active repeating FRBs, namely FRB~20121102A, FBR~20190520B, and FRB~20201124A, do not appear to be identical. This suggests that repeating FRBs may have diverse emission properties.
    
\end{itemize}

\normalem
\begin{acknowledgements}
This work made use of the data from FAST, a Chinese national mega-science facility, operated by National Astronomical Observatories, Chinese Academy of Sciences. This work supported by the Open Project Program of the Key Laboratory of FAST, Chinese Academy of Sciences. We acknowledge use of the CHIME/FRB Public Database, provided at https://www.chime-frb.ca/ by the CHIME/FRB Collaboration. P. W. acknowledges support from the National Natural Science Foundation of China under grant U2031117, the Youth Innovation Promotion Association CAS (id. 2021055), CAS Project for Young Scientists in Basic Reasearch (grant YSBR-006) and the Cultivation Project for FAST Scientific Payoff and Research Achievement of CAMS-CAS. Y. Feng is supported by the Key Research Project of Zhejiang Lab no. 2021PE0AC0. J.~L. Han is supported by the National Natural Science Foundation of China (NSFC, Nos. 11988101 and 11833009) and the Key Research Program of the Chinese Academy of Sciences (Grant No. QYZDJ-SSW-SLH021); D.~J. Zhou is supported by the Cultivation Project for the FAST scientific Payoff and Research Achievement of CAMS-CAS. K.J.~Lee, J.C.~Jiang, H.~Xu, and J.W.~Xu are supported by National SKA Program of China 2020SKA0120200, National Nature Science Foundation grant No. 12041303, the CAS-MPG LEGACY project, and funding from the Max-Planck Partner Group.

\end{acknowledgements}

\section*{Author Contributions}
Y.K.Z., P.W., D.L., and B.Z. developed the concept of the manuscript. B.Z. and W.W.Z. proposed and chaired the FAST FRB key science project. Y.K.Z. conducted the data analysis and visualization. Y.K.Z., P.W., and C.H.N. searched the bursts and analyzed the burst properties. Y.K.Z., Y.F., D.L., B.Z., C.W.T., and R.L. led the discussion on the interpretation of the results and writing of the manuscript. D.J.Z. studied bursts morphology and classification with the details presented in paper I of this series. J.C.J performed the analysis of the polarization properties and the results are presented in the Paper III. J.R.N. performed periodicity search with the results presented in paper IV. All authors contributed to the analysis or interpretation of the data and to the final version of the manuscript.

\bibliographystyle{raa}
\bibliography{bibtex}

\begin{thebibliography}{65}
\providecommand\natexlab[1]{#1}
\providecommand\JournalTitle[1]{#1}

\bibitem[{Aggarwal}(2021)]{2021ApJ...920L..18A}
{Aggarwal}, K. 2021, \apjl, 920, L18

\bibitem[{Aggarwal} {et~al.}(2021)]{2021ApJ...922..115A}
{Aggarwal}, K., {Agarwal}, D., {Lewis}, E.~F., {et~al.} 2021, \apj, 922, 115

\bibitem[{Arcus} {et~al.}(2021)]{2021MNRAS.501.5319A}
{Arcus}, W.~R., {Macquart}, J.~P., {Sammons}, M.~W., {James}, C.~W., \&
  {Ekers}, R.~D. 2021, \mnras, 501, 5319

\bibitem[{Band} {et~al.}(1993)]{1993ApJ...413..281B}
{Band}, D., {Matteson}, J., {Ford}, L., {et~al.} 1993, \apj, 413, 281

\bibitem[{Bhandari} {et~al.}(2018)]{2018MNRAS.475.1427B}
{Bhandari}, S., {Keane}, E.~F., {Barr}, E.~D., {et~al.} 2018, \mnras, 475, 1427

\bibitem[{Chatterjee} {et~al.}(2017)]{2017Natur.541...58C}
{Chatterjee}, S., {Law}, C.~J., {Wharton}, R.~S., {et~al.} 2017, \nat, 541, 58

\bibitem[{CHIME/FRB Collaboration} {et~al.}(2020)]{2020Natur.582..351C}
{CHIME/FRB Collaboration}, {Amiri}, M., {Andersen}, B.~C., {et~al.} 2020, \nat,
  582, 351

\bibitem[{CHIME/FRB Collaboration} {et~al.}(2021)]{2021ApJS..257...59C}
{CHIME/FRB Collaboration}, {Amiri}, M., {Andersen}, B.~C., {et~al.} 2021,
  \apjs, 257, 59

\bibitem[{CHIME/FRB Collabortion}(2021)]{2021ATel14497....1C}
{CHIME/FRB Collabortion}. 2021, The Astronomer's Telegram, 14497, 1

\bibitem[{Cruces} {et~al.}(2021)]{2021MNRAS.500..448C}
{Cruces}, M., {Spitler}, L.~G., {Scholz}, P., {et~al.} 2021, \mnras, 500, 448

\bibitem[{Fong} {et~al.}(2021)]{2021ApJ...919L..23F}
{Fong}, W.-f., {Dong}, Y., {Leja}, J., {et~al.} 2021, \apjl, 919, L23

\bibitem[{Golpayegani} {et~al.}(2019)]{2019MNRAS.489.4001G}
{Golpayegani}, G., {Lorimer}, D.~R., {Ellingson}, S.~W., {et~al.} 2019, \mnras,
  489, 4001

\bibitem[{Gourdji} {et~al.}(2019)]{2019ApJ...877L..19G}
{Gourdji}, K., {Michilli}, D., {Spitler}, L.~G., {et~al.} 2019, \apjl, 877, L19

\bibitem[{Hashimoto} {et~al.}(2020{\natexlab{a}})]{2020MNRAS.494.2886H}
{Hashimoto}, T., {Goto}, T., {Wang}, T.-W., {et~al.} 2020{\natexlab{a}},
  \mnras, 494, 2886

\bibitem[{Hashimoto} {et~al.}(2020{\natexlab{b}})]{2020MNRAS.498.3927H}
{Hashimoto}, T., {Goto}, T., {On}, A. Y.~L., {et~al.} 2020{\natexlab{b}},
  \mnras, 498, 3927

\bibitem[{Hashimoto} {et~al.}(2022)]{2022MNRAS.511.1961H}
{Hashimoto}, T., {Goto}, T., {Chen}, B.~H., {et~al.} 2022, \mnras, 511, 1961

\bibitem[{Hewitt} {et~al.}(2022)]{2022MNRAS.515.3577H}
{Hewitt}, D.~M., {Snelders}, M.~P., {Hessels}, J.~W.~T., {et~al.} 2022, \mnras,
  515, 3577

\bibitem[{Hilmarsson} {et~al.}(2021)]{2021MNRAS.508.5354H}
{Hilmarsson}, G.~H., {Spitler}, L.~G., {Main}, R.~A., \& {Li}, D.~Z. 2021,
  \mnras, 508, 5354

\bibitem[{James} {et~al.}(2022)]{2022MNRAS.510L..18J}
{James}, C.~W., {Prochaska}, J.~X., {Macquart}, J.~P., {et~al.} 2022, \mnras,
  510, L18

\bibitem[{Jiang} {et~al.}(2020)]{2020RAA....20...64J}
{Jiang}, P., {Tang}, N.-Y., {Hou}, L.-G., {et~al.} 2020, Research in Astronomy
  and Astrophysics, 20, 064

\bibitem[{Kumar} {et~al.}(2017)]{2017MNRAS.468.2726K}
{Kumar}, P., {Lu}, W., \& {Bhattacharya}, M. 2017, \mnras, 468, 2726

\bibitem[{Kumar} {et~al.}(2022)]{2022MNRAS.512.3400K}
{Kumar}, P., {Shannon}, R.~M., {Lower}, M.~E., {et~al.} 2022, \mnras, 512, 3400

\bibitem[{Lanman} {et~al.}(2022)]{2022ApJ...927...59L}
{Lanman}, A.~E., {Andersen}, B.~C., {Chawla}, P., {et~al.} 2022, \apj, 927, 59

\bibitem[{Law} {et~al.}(2017)]{2017ApJ...850...76L}
{Law}, C.~J., {Abruzzo}, M.~W., {Bassa}, C.~G., {et~al.} 2017, \apj, 850, 76

\bibitem[{Law} {et~al.}(2021)]{2021ATel14526....1L}
{Law}, C., {Tendulkar}, S., {Clarke}, T., {Aggarwal}, K., \& {Bethapudy}, S.
  2021, The Astronomer's Telegram, 14526, 1

\bibitem[Li {et~al.}(2019)]{Li_2019}
Li, D., Dickey, J.~M., \& Liu, S. 2019, Research in Astronomy and Astrophysics,
  19, 016

\bibitem[{Li} {et~al.}(2018)]{FLAN}
{Li}, D., {Wang}, P., {Qian}, L., {et~al.} 2018, IEEE Microwave Magazine, 19,
  112

\bibitem[{Li} {et~al.}(2021)]{2021Natur.598..267L}
{Li}, D., {Wang}, P., {Zhu}, W.~W., {et~al.} 2021, \nat, 598, 267

\bibitem[{Lorimer} {et~al.}(2007)]{2007Sci...318..777L}
{Lorimer}, D.~R., {Bailes}, M., {McLaughlin}, M.~A., {Narkevic}, D.~J., \&
  {Crawford}, F. 2007, Science, 318, 777

\bibitem[{Lu} \& {Kumar}(2016)]{2016MNRAS.461L.122L}
{Lu}, W., \& {Kumar}, P. 2016, \mnras, 461, L122

\bibitem[{Lu} \& {Kumar}(2019)]{2019MNRAS.483L..93L}
{Lu}, W., \& {Kumar}, P. 2019, \mnras, 483, L93

\bibitem[{Lu} {et~al.}(2020)]{lu20}
{Lu}, W., {Kumar}, P., \& {Zhang}, B. 2020, \mnras, 498, 1397

\bibitem[{Luo} {et~al.}(2018)]{2018MNRAS.481.2320L}
{Luo}, R., {Lee}, K., {Lorimer}, D.~R., \& {Zhang}, B. 2018, \mnras, 481, 2320

\bibitem[{Luo} {et~al.}(2020)]{2020MNRAS.494..665L}
{Luo}, R., {Men}, Y., {Lee}, K., {et~al.} 2020, \mnras, 494, 665

\bibitem[{Lyu} {et~al.}(2021)]{2021FrPhy..1624503L}
{Lyu}, F., {Meng}, Y.-Z., {Tang}, Z.-F., {et~al.} 2021, Frontiers of Physics,
  16, 24503

\bibitem[{Macquart} \& {Ekers}(2018)]{2018MNRAS.474.1900M}
{Macquart}, J.~P., \& {Ekers}, R.~D. 2018, \mnras, 474, 1900

\bibitem[{Madau} \& {Dickinson}(2014)]{2014ARA&A..52..415M}
{Madau}, P., \& {Dickinson}, M. 2014, \araa, 52, 415

\bibitem[{Maehara} {et~al.}(2015)]{2015EP&S...67...59M}
{Maehara}, H., {Shibayama}, T., {Notsu}, Y., {et~al.} 2015, Earth, Planets and
  Space, 67, 59

\bibitem[{Main} {et~al.}(2021)]{2021MNRAS.tmp.2928M}
{Main}, R.~A., {Hilmarsson}, G.~H., {Marthi}, V.~R., {et~al.} 2021,
  arXiv:2108.00052

\bibitem[{Marcote} {et~al.}(2017)]{2017ApJ...834L...8M}
{Marcote}, B., {Paragi}, Z., {Hessels}, J.~W.~T., {et~al.} 2017, \apjl, 834, L8

\bibitem[{Marthi} {et~al.}(2021)]{2021MNRAS.tmp.2782M}
{Marthi}, V.~R., {Bethapudi}, S., {Main}, R.~A., {et~al.} 2021,
  arXiv:2108.00697

\bibitem[{Mickaliger} {et~al.}(2012)]{2012ApJ...760...64M}
{Mickaliger}, M.~B., {McLaughlin}, M.~A., {Lorimer}, D.~R., {et~al.} 2012,
  \apj, 760, 64

\bibitem[{Nan} {et~al.}(2011)]{nan11}
{Nan}, R., {Li}, D., {Jin}, C., {et~al.} 2011, International Journal of Modern
  Physics D, 20, 989

\bibitem[{Nicholl} {et~al.}(2017)]{2017ApJ...843...84N}
{Nicholl}, M., {Williams}, P.~K.~G., {Berger}, E., {et~al.} 2017, \apj, 843, 84

\bibitem[{Nimmo} {et~al.}(2022)]{2022ApJ...927L...3N}
{Nimmo}, K., {Hewitt}, D.~M., {Hessels}, J.~W.~T., {et~al.} 2022, \apjl, 927,
  L3

\bibitem[{Niu} {et~al.}(2022)]{2022Natur.606..873N}
{Niu}, C.~H., {Aggarwal}, K., {Li}, D., {et~al.} 2022, \nat, 606, 873

\bibitem[{Petroff} {et~al.}(2016)]{2016PASA...33...45P}
{Petroff}, E., {Barr}, E.~D., {Jameson}, A., {et~al.} 2016, \pasa, 33, e045

\bibitem[{Piro} {et~al.}(2021{\natexlab{a}})]{2021A&A...656L..15P}
{Piro}, L., {Bruni}, G., {Troja}, E., {et~al.} 2021{\natexlab{a}}, \aap, 656,
  L15

\bibitem[{Piro} {et~al.}(2021{\natexlab{b}})]{piro21}
{Piro}, L., {Bruni}, G., {Troja}, E., {et~al.} 2021{\natexlab{b}}, \aap, 656,
  L15

\bibitem[{Pleunis} {et~al.}(2021)]{2021ApJ...923....1P}
{Pleunis}, Z., {Good}, D.~C., {Kaspi}, V.~M., {et~al.} 2021, \apj, 923, 1

\bibitem[{Ransom}(2001)]{2001PhDT.......123R}
{Ransom}, S.~M. 2001, {New search techniques for binary pulsars}, PhD thesis,
  Harvard University, Massachusetts

\bibitem[{Ravi} {et~al.}(2022)]{2022MNRAS.513..982R}
{Ravi}, V., {Law}, C.~J., {Li}, D., {et~al.} 2022, \mnras, 513, 982

\bibitem[{Schechter}(1976)]{1976ApJ...203..297S}
{Schechter}, P. 1976, \apj, 203, 297

\bibitem[{Spitler} {et~al.}(2014)]{2014ApJ...790..101S}
{Spitler}, L.~G., {Cordes}, J.~M., {Hessels}, J.~W.~T., {et~al.} 2014, \apj,
  790, 101

\bibitem[{Spitler} {et~al.}(2016)]{2016Natur.531..202S}
{Spitler}, L.~G., {Scholz}, P., {Hessels}, J.~W.~T., {et~al.} 2016, \nat, 531,
  202

\bibitem[{Sridhar} {et~al.}(2021)]{2021ApJ...917...13S}
{Sridhar}, N., {Metzger}, B.~D., {Beniamini}, P., {et~al.} 2021, \apj, 917, 13

\bibitem[{Vedantham} {et~al.}(2016)]{2016ApJ...830...75V}
{Vedantham}, H.~K., {Ravi}, V., {Hallinan}, G., \& {Shannon}, R.~M. 2016, \apj,
  830, 75

\bibitem[{Wang} \& {Zhang}(2019)]{2019ApJ...882..108W}
{Wang}, F.~Y., \& {Zhang}, G.~Q. 2019, \apj, 882, 108

\bibitem[{Xu} {et~al.}(2021)]{2021arXiv211111764X}
{Xu}, H., {Niu}, J.~R., {Chen}, P., {et~al.} 2021, arXiv e-prints,
  arXiv:2111.11764

\bibitem[{Yang} \& {Zhang}(2018)]{2018ApJ...868...31Y}
{Yang}, Y.-P., \& {Zhang}, B. 2018, \apj, 868, 31

\bibitem[{Zhang}(2018)]{2018ApJ...867L..21Z}
{Zhang}, B. 2018, \apjl, 867, L21

\bibitem[{Zhang}(2020)]{zhang20}
{Zhang}, B. 2020, \nat, 587, 45

\bibitem[{Zhang}(2022)]{2022ApJ...925...53Z}
{Zhang}, B. 2022, \apj, 925, 53

\bibitem[{Zhang} \& {Zhang}(2022)]{2022ApJ...924L..14Z}
{Zhang}, R.~C., \& {Zhang}, B. 2022, \apjl, 924, L14

\bibitem[{Zhang} {et~al.}(2021)]{2021MNRAS.501..157Z}
{Zhang}, R.~C., {Zhang}, B., {Li}, Y., \& {Lorimer}, D.~R. 2021, \mnras, 501,
  157

\end{thebibliography}

\begin{appendix}
\onecolumn
\setlength{\tabcolsep}{5pt}
\begin{landscape}
\section{Bursts Properties}

\begin{center}
\begin{ThreePartTable}
\begin{TableNotes}
  \item [a] {At 1.5 GHz}
  \item [b] {Equivalent width}
  \item [c] {Energy calculated with center frequency.}
  \item [d] {Energy calculated with band width.}
\end{TableNotes}
\begin{tiny}
\begin{longtable}{ccccccccccccccc}
  \rmfamily\\
  \caption{The properties of the FRB~20201124A bursts}\label{tab:burst}\\
  \toprule
 Burst ID &  ObsDate & MJD\tnote{a} & DM             &  Width\tnote{b} & PeakFlux  & Fluence   & Energy$_{c}$\tnote{c} & Energy$_{w}$\tnote{d} & BandWidth & Width\tnote{b} & PeakFlux     & Fluence      & Energy$_{w}$\tnote{c} \\
          &          & Barycentric  &                &  Full Band      & Full Band & Full Band & Full Band             & Full Band             &           & Limited Band   & Limited Band & Limited Band & Limited Band          \\
          &          &              & (pc cm$^{-3}$) &  (ms)           & (mJy)     & (Jy ms)   & (erg)                 & (erg)                 &  (MHz)    & (ms)           & (mJy)        & (Jy ms)      & (erg)                 \\
  \midrule
  \insertTableNotes
  \endfirsthead

  \toprule
  \endhead

  \bottomrule
  \endfoot

  \bottomrule
  \endlastfoot

\midrule
     B001 &  20210925 &  59482.944514719 & 412.97 &   3.31 &   12.39$\pm$0.26 &    0.04$\pm$0.86 &    1.148(1)e+37 &    4.590(1)e+36 &     200 &  3.81 &    22.50$\pm$0.74 &   0.09$\pm$0.01 &   4.210(45)e+36 \\
     B002 &  20210925 &  59482.946731974 & 412.97 &   3.78 &   28.61$\pm$0.05 &     0.11$\pm$0.2 &   3.023(28)e+37 &   1.209(11)e+37 &     150 &  3.55 &    79.38$\pm$0.36 &   0.28$\pm$0.01 &   1.038(10)e+37 \\
     B003 &  20210925 &  59482.947007613 & 412.97 &   4.34 &    5.99$\pm$0.07 &    0.03$\pm$0.29 &    7.278(1)e+36 &    2.911(1)e+36 &      50 &  5.18 &    32.52$\pm$0.22 &   0.17$\pm$0.01 &   2.071(22)e+36 \\
     B004 &  20210925 &  59482.947121260 & 412.97 &   2.70 &    7.41$\pm$0.09 &    0.02$\pm$0.24 &    5.598(1)e+36 &    2.239(1)e+36 &     100 &  3.54 &    20.05$\pm$0.42 &   0.07$\pm$0.01 &   1.747(22)e+36 \\
     B005 &  20210925 &  59482.947355910 & 412.97 &   4.47 &     9.61$\pm$0.1 &    0.04$\pm$0.45 &    1.204(1)e+37 &    4.814(1)e+36 &     100 &  4.00 &    38.02$\pm$0.22 &   0.15$\pm$0.01 &   3.740(45)e+36 \\
     B006 &  20210925 &  59482.947576756 & 412.97 &   2.85 &   10.89$\pm$1.93 &    0.03$\pm$5.49 &    8.677(1)e+36 &    3.471(1)e+36 &     150 &  3.17 &    22.96$\pm$4.26 &   0.07$\pm$0.01 &   2.687(34)e+36 \\
     B007 &  20210925 &  59482.947690227 & 412.97 &   8.46 &   211.11$\pm$1.1 &    1.79$\pm$9.33 &   5.002(45)e+38 &   2.001(18)e+38 &     300 &  7.82 &   462.35$\pm$2.39 &   3.62$\pm$0.03 &   2.668(25)e+38 \\
     B008 &  20210925 &  59482.950048429 & 412.97 &   5.77 &   68.47$\pm$0.64 &     0.40$\pm$3.7 &   1.106(11)e+38 &   4.422(45)e+37 &     450 &  5.26 &    75.19$\pm$0.71 &   0.40$\pm$0.01 &   4.373(40)e+37 \\
\bottomrule
\end{longtable}
\end{tiny}
\end{ThreePartTable}
\end{center}
\end{landscape}

\twocolumn
\end{appendix}

\end{document}